\documentclass[journal,twoside,web]{IEEEtran}

\usepackage{graphicx}
\usepackage{cite}
\usepackage{amsmath,amssymb,amsfonts}
\usepackage{algorithm}
\usepackage{algorithmic}
\usepackage{textcomp}
\usepackage{booktabs}
\usepackage{color,soul}
\usepackage{xspace}
\usepackage[colorlinks]{hyperref}
\usepackage{amsmath}
\usepackage{footnote}
\usepackage{makecell}
\usepackage{multirow}
\usepackage[export]{adjustbox}

\usepackage{array}
\usepackage[nolist,nohyperlinks]{acronym}
\usepackage{textcomp}

\usepackage{titlesec}
\def\BibTeX{{\rm B\kern-.05em{\sc i\kern-.025em b}\kern-.08em
    T\kern-.1667em\lower.7ex\hbox{E}\kern-.125emX}}
\makesavenoteenv{tabular}

\DeclareMathOperator{\CLR}{CLR}
\DeclareMathOperator{\conv}{Conv}

\DeclareMathOperator{\DSDF}{DSDF}

\newcommand{\loss}{\ensuremath{\mathcal{L}}}
\newcommand{\lbce}{\ensuremath{\loss_\text{BCE}}\xspace}
\newcommand{\ldcs}{\ensuremath{\loss_\text{DCS}}\xspace}
\newcommand{\lcomb}{\ensuremath{\loss_\text{comb}}\xspace}
\newcommand{\lmsrf}{\ensuremath{\loss_\text{MSRF}}\xspace}

\newcommand{\sysname}{\text{MSRF-Net}\xspace}
\usepackage[acronym]{glossaries}

\acrodef{CT}{Computed Tomography}
\acrodef{CNN}{Convolutional Neural Network}
\acrodef{ASPP}{Atrous Spatial Pyramid Pooling}
\acrodef{mIoU}{mean Intersection over Union}
\acrodef{DSC}{Dice Coefficient}
\acrodef{FCN}{Fully Convolutional Network}
\acrodef{SOTA}{State Of The Art}
\acrodef{DSDF}{Dual-Scale Dense Fusion}
\acrodef{MSRF}{Multi-Scale Residual Fusion}
\acrodef{DSB}{Data Science Bowl}
\acrodef{MRI}{Magnetic Resonance Imaging}
\acrodef{PPM}{Pyramid Pooling Module}
\acrodef{FPS}{Frames Per Second}

\begin{document}
\title{MSRF-Net: A Multi-Scale Residual Fusion Network  for Biomedical Image Segmentation}
\author{Abhishek Srivastava,
        Debesh Jha, \IEEEmembership{Member, IEEE}, 
       Sukalpa Chanda, \IEEEmembership{Member, IEEE},
       Umapada Pal, \IEEEmembership{Senior Member, IEEE},
       H{\aa}vard D. Johansen, \IEEEmembership{Member, IEEE},
       Dag Johansen, \IEEEmembership{Member, IEEE}, \\
       Michael A. Riegler,  \IEEEmembership{Member, IEEE},
       Sharib Ali, \IEEEmembership{Member, IEEE},
       P{\aa}l Halvorsen \IEEEmembership{Member, IEEE}
\thanks{A. Srivastava is with  Computer Vision and Pattern Recognition Unit, Indian Statistical Institute, Kolkata, India}
\thanks{D. Jha is with SimulaMet, Oslo, Norway and UiT The Arctic University of Norway, Troms{\o}, Norway (corresponding email: debesh@simula.no)}
\thanks{S. Chanda is with Østfold University College, Halden, Norway}
\thanks{U. Pal is with Indian Statistical Institute, Kolkata, India}
\thanks{H. D. Johansen and D. Johansen are with UiT The Arctic University of Norway, Troms{\o}, Norway}
\thanks{M. A. Riegler is with SimulaMet, Oslo, Norway and UiT The Arctic University of Norway, Troms{\o}, Norway}
\thanks{S. Ali is with the Department of Engineering Science, University of Oxford, and Oxford NIHR Biomedical Research Centre, Oxford, UK (corresponding email: sharib.ali@eng.ox.ac.uk)}
\thanks{P. Halvorsen is with SimulaMet, Oslo, Norway and Oslo Metropolitan University, Oslo, Norway}
\thanks{S. Ali and P. Halvorsen: Shared senior authorship} 
}
\markboth{
}{Srivastava \MakeLowercase{\textit{et al.}}: MSRF-Net: A Multi-Scale Residual Fusion Network  for Biomedical Image Segmentation}

\maketitle
\begin{abstract}
Methods based on convolutional neural networks have improved the performance of biomedical image segmentation. However, most of these methods cannot efficiently segment objects of variable sizes and train on small and biased datasets, which are common for biomedical use cases. While methods exist that incorporate multi-scale fusion approaches to address the challenges arising with variable sizes, they usually use complex models that are more suitable for general semantic segmentation problems. In this paper, we propose a novel architecture called Multi-Scale Residual Fusion Network (\sysname), which is specially designed for medical image segmentation. The proposed \sysname is able to exchange multi-scale features of varying receptive fields using a Dual-Scale Dense Fusion (DSDF) block. Our DSDF block can exchange information rigorously across two different resolution scales, and our MSRF sub-network uses multiple DSDF blocks in sequence to perform multi-scale fusion. This allows the preservation of resolution, improved information flow and propagation of both high- and low-level features to obtain accurate segmentation maps. The proposed \sysname allows to capture object variabilities and provides improved results on different biomedical datasets. Extensive experiments on \sysname demonstrate that the proposed method outperforms the cutting-edge medical image segmentation methods on four publicly available datasets. We achieve the \ac{DSC} of 0.9217, 0.9420, and 0.9224, 0.8824 on Kvasir-SEG, CVC-ClinicDB, 2018 Data Science Bowl dataset, and ISIC-2018 skin lesion segmentation challenge dataset respectively. We further conducted generalizability tests and achieved \ac{DSC} of 0.7921 and 0.7575 on  CVC-ClinicDB and Kvasir-SEG, respectively.

\end{abstract}

\begin{IEEEkeywords}
Medical image segmentation, MSRF-Net, multi-scale fusion, generalization, colonoscopy 
\end{IEEEkeywords}

\IEEEpeerreviewmaketitle

\section{Introduction}
\label{sec:introduction}
\IEEEPARstart{M}EDICAL image segmentation is an essential task in clinical diagnosis and has been extensively studied by the medical image analysis community~\cite{celebi2019dermoscopy,caicedo2019nucleus,ali2021deep}. The semantic segmentation results can help identify regions of interest for lesion assessment, such as polyps in the colon, to inspect if they are cancerous and remove them if necessary. Thus, the segmentation results can help to detect missed lesions, prevent diseases, and improve therapy planning and treatment. The significant challenge in medical imaging is the requirement of a large number of high-quality labeled and annotated datasets. This is a key factor in the development of robust algorithm for automated medical image segmentation task.   

The manual pixel-wise annotation of medical image data is very time-consuming, requires collaborations with experienced medical experts, and is costly. During the annotation of the regions in medical images (for example, polyps in still frames), the guidelines and protocols are set based on which expert performs the annotations. However, there might exist discrepancies among the experts, e.g., while considering a particular area in the lesion as cancerous or non-cancerous. Additionally, the lack of standard annotation protocols for various imaging modalities and low image quality can influence annotation quality. Other factors such as the annotator's attentiveness, type of display device, image-annotation software and data misinterpretation due to lightning conditions can also affect the quality of annotations \cite{lux2013annotation}. An alternative solution to manual image segmentation is an automated computer aided segmentation based diagnosis-assisting system that can provide a faster, more accurate, and more reliable solution to transform clinical procedures and improve patient care. Computer aided diagnosis will reduce the expert's burden and also reduce the overall treatment cost. Due to the diverse nature of medical-imaging data, computer aided diagnosis based segmentation models must be robust to variations in imaging modalities \cite{jha2021real}. 

In the past years, \acp{CNN} based approaches have overcome the limitations of traditional segmentation methods~\cite{litjens2017survey} in various medical imaging modalities such as X-ray, \ac{CT}, \ac{MRI}, endoscopy, wireless capsule endoscopy, dermatoscopy, and in high-throughput imaging like histopathology and electron microscopy. Modern semantic and instance segmentation architectures are usually encoder-decoder based networks~\cite{shen2017deep,ross2020robust}. The success of deep encoder-decoder based \acp{CNN} is largely due to their skip connections, which allows the propagation of deep, semantically meaningful, and dense feature maps from the encoder network to the decoder sub-networks~\cite{drozdzal2016importance,jha2020doubleu}. However, encoder-decoder based image segmentation architectures have limitations in optimal depth and design of the skip connections~\cite{zhou2019unet++}. The optimal depth of the architectures can vary from one biomedical application to another. The number of samples in the dataset used in training also contributes to the limitation on the complexity of the network. The design of skip connections are sometimes unnecessarily restrictive, demanding the fusion of the same-scale encoder and decoder feature maps. Moreover, traditional \ac{CNN} methods do not make use of the hierarchical features.

In this paper, we propose a novel medical image segmentation architecture, called \emph{\sysname}, which aims to overcome the above discussed limitations. Our proposed \sysname maintains high-resolution representation throughout the process, which is conducive to potentially achieving high spatial accuracy. The \sysname utilizes a novel \ac{DSDF} block that performs dual scale feature exchange and a sub-network that exchanges multi-scale features using the \ac{DSDF} block. The \ac{DSDF} block takes two different scale inputs and employs a residual dense block that exchanges information across different scales after each convolutional layer in their corresponding dense blocks. The densely connected nature of blocks allows relevant high- and low-level features to be preserved for the final segmentation map prediction. The multi-scale information exchange in our network preserves both high- and low-resolution feature representations, thereby producing finer, richer and spatially accurate segmentation maps. The repeated multi-scale fusion helps in enhancing the high-resolution feature representations with the information propagated by low-resolution representations. Further, layers of residual networks allow redundant \ac{DSDF} blocks to die out, and only the most relevant extracted features contribute to the predicted segmentation maps. 

Additionally, we propose adding a complimentary gated shape stream that can leverage the combination of high- and low-level features to compute shape boundaries accurately. 
We have evaluated the \sysname segmentation model using four publicly available biomedical datasets. The results demonstrate that the proposed \sysname outperforms the \ac{SOTA} segmentation methods on most standard computer vision evaluation metrics.

The main contributions of this work are as following:
\begin{enumerate}
\item Our proposed \sysname architecture is based on a \ac{DSDF} block that comprises of residual dense connections and exchanging information across multiple scales. This allows both high-resolution and low-resolution features to propagate, thereby extracting semantically meaningful features that improve segmentation performance on various biomedical datasets. 

\item \sysname computes the multi-scale features and fuses them effectively using a \ac{DSDF} block. The residual nature of \ac{DSDF} block improves gradient flow which improves the training efficiency, i.e., reducing the need for large datasets for training.    

\item The effectiveness of \sysname is demonstrated on four public datasets: Kvasir-SEG~\cite{jha2020kvasir}, CVC-ClinicDB~\cite{bernal2015wm}, 2018 \acf{DSB} Challenge~\cite{caicedo2019nucleus}, and ISIC 2018 Challenge~\cite{codella2018skin,tschandl2018ham10000}. We conduct a generalizability study of the proposed network for which we trained our model on Kvasir-SEG and tested on the CVC-ClinicDB and vice versa. The experimental results and their comparison with established computer vision methods confirmed that our approach is more generalizable.

\end{enumerate}

\section{Related work}

\subsection{Medical image segmentation}
Long et al.~\cite{long2015fully} proposed a \ac{FCN} that included only convolutional layers for semantic segmentation. Subsequently, Ronneberger et al.~\cite{ronneberger2015u} modified the \ac{FCN} with an encoder-decoder U-Net architecture for segmentation of HeLa cells and neuronal structures of electron microscopic stacks. In the U-Net~\cite{ronneberger2015u}, low- and high-level feature maps are combined through skip connections. The high-level feature maps are processed by deeper layers of the encoder network and propagated through the decoder whereas, the low-level features are propagated from the initial layers of the network. This may cause a semantic gap between the high- and low-level features. Ibtehaz et al.~\cite{ibtehaz2020multiresunet} proposed to add convolutional units along the path of skip connections to reduce the semantic gap. Oktay et al.~\cite{oktay2018attention} proposed an attention U-Net that used an attention block to alter the feature maps propagated through the skip-connections. Here, the previous decoder block output was used to form a gating mechanism to prune unnecessary spatial features passing from the skip-connections and to keep only the relevant features. In addition, various other extensions of the U-Net have been proposed~\cite{zhou2019unet++,jha2019resunet++,jha2020doubleu,zhou2018unet++,sun2020saunet}. To incorporate global context information for the task of scene parsing, PSPNet~\cite{zhao2017pyramid} generated hierarchical feature maps through a \ac{PPM}. Similarly, Chen et al.~\cite{chen2017deeplab} used the \ac{ASPP} to aggregate the global features. Later, the same group proposed the DeepLabV3+~\cite{chen2018encoder} architecture that used skip connections between the encoder and decoder. Both of these networks have been widely used by the biomedical imaging community~\cite{hassan2020,rad2019cell,guo2020polyp}.

Hu et al.~\cite{hu2018squeeze} proposed SE-Net, which pioneered channel-wise attention. The Squeeze and Excitation ($\text{S\&E}$) block was able to model interdependencies between the channels and derive a global information map that helps in emphasizing relevant features and suppressing irrelevant features. FED-Net~\cite{chen2019feature} incorporated these $\text{S\&E}$ blocks in their modified U-Net architecture. Kaul et al.~\cite{kaul2019focusnet} incorporated both types of attention, i.e., spatial and channel-wise attention, in their proposed FocusNet. Jha et al.~\cite{jha2019resunet++} modified ResUNet~\cite{zhang2018road} adding \ac{ASPP}, $\text{S\&E}$ block~\cite{hu2018squeeze} and attention mechanisms to boost the performance of the network further. Taikkawa et al.~\cite{takikawa2019gated} proposed Gated-SCNN, which pioneered the idea of gated shape stream to generate finer segmentation maps leveraging the shape and boundaries of the target object. The shape stream was recently also employed by Sun at al.~\cite{sun2020saunet} to capture the shape and boundaries of the target segmentation map for medical segmentation problems. Fan et al.~\cite{fan2020pranet} devised a parallel partial decoder (PraNet) that aggregated high-level features to generate a guidance map that estimates a rough location of the region of interest. The guidance map used in PraNet was then used with a reverse attention module to extract finer boundaries from the low-level features. Kim et al.~\cite{kim2021uacanet} modified the U-Net architecture and added additional encoder and decoder modules. Saliency maps computed by a prediction module in the UACA-Net is used to compute foreground, background and uncertain area maps for each representation. The relationship between each representation is computed and used by the next prediction module. A detailed summary of advances of deep-learning based methodologies in medical image segmentation can be found in~\cite{hesamian2019deep,sarvamangala2021convolutional,liu2020survey}.  

\begin{figure}[t!]
\includegraphics[width=0.5\textwidth]{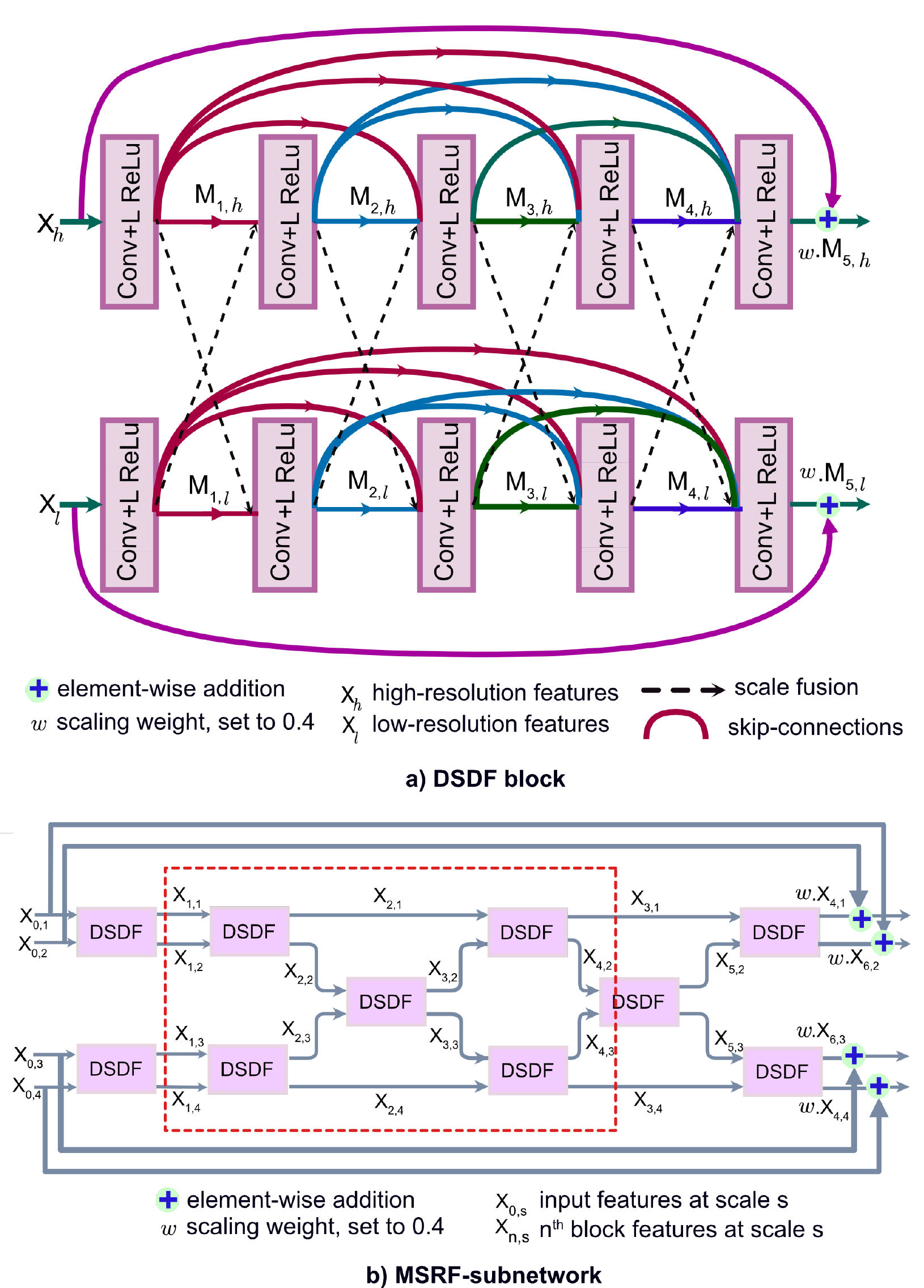}
\caption{Components of our MSRF-Net, a) Proposed \acf{DSDF} block and b) \acf{MSRF}. Dotted rectangle block in (b) represents multi-scale feature exchange in MSRF-Net}
\label{fig:smfrdbblock}
\vspace{-5mm}
\end{figure}
\subsection{Residual dense blocks}
Dense connections are a unique approach of improving information flow and keeping a collection of diversified features.
The architectures based on dense connections are characterized by each layer receiving inputs from all previous layers.
Various medical image segmentation methods~\cite{guan2019fully,zhou2019unet++,zhang2018residual,yang2019road,ding2019deep} leverage the diversified features captured by such dense connections to improve segmentation performance. Guan et al.~\cite{guan2019fully} modified the U-Net architecture by substituting standard encoder-decoder units with densely connected convolutional units. Zhou et al.~\cite{zhou2019unet++} conceived an architecture where the encoder and decoder are connected through dense and nested skip pathways for efficient feature fusion between the feature maps of encoder and decoder. Zhang et al.~\cite{zhang2018residual} proposed Residual Dense Blocks (RDB) to extract local features via densely connected convolutional layers. Additionally, their architecture allowed them to connect the previous RDB block to all the current RDB blocks and a final global fusion through $1\times 1$ convolutions for maintaining global hierarchical feature extraction. In ResUNet~\cite{yang2019road} and Residual Dense U-Net (RD-U-Net)~\cite{ding2019deep}, the RDBs are included in a standard U-Net based architecture to make use of hierarchical features. Dolz el al.~\cite{dolz2018hyperdense} proposed HyperDense-Net, which introduced a two-stream \ac{CNN} designed to process each modality in a separate stream for multi-modal image segmentation. The dense connections were used across layers of the same path and also between layers of a different path, therefore, increasing the capacity of the network to learn more complex combination between different modality.

\subsection{Multi-scale fusion}
Maintaining a high-resolution representation of the image is important for segmentation architectures to precisely capture the spatial information and give accurate segmentation maps~\cite{Wang_2020}.
Rather than recovering such representations from low-level representations, multi-scale fusion can help exchange high- and low-resolution features throughout the segmentation process. Wang et al.~\cite{Wang_2020} demonstrated that such exchange of features improves the flow of high-resolution features and can potentially lead to a more spatially accurate segmentation map. They achieved this by processing all the resolution streams in parallel, keeping the resolution representation for each resolution, and performing the feature fusion across all resolution scales.

The previous works by Ronneberger et al.~\cite{ronneberger2015u} and Badrinarayanan et al.~\cite{badrinarayanan2017segnet} used skip-connections to concatenate high-resolution feature representations at each level with the upscaled features in the decoder to preserve both high- and low-resolution feature representations. Zhao et al.~\cite{zhao2017pyramid} used pyramid pooling to perform multi-resolution fusion while Chen et al.~\cite{chen2017deeplab} used \ac{ASPP} and multiple Atrous convolutions with different sampling rates. Similarly, Yang et al.~\cite{yang2018denseaspp} used densely connected atrous convolutional layers in their DenseASPP network to gather multi-scale features with a large range of receptive fields. Lin et al.~\cite{lin2019zigzagnet} proposed ZigZagNet, which fused multi-resolution features by exchanging information in a zig-zag fashion between the encoder-decoder architecture. Wang et al.~\cite{wang2016deeply} proposed Deeply-Fused Nets that applies fusion of intermediate resolutions allowing varying receptive fields with different sizes. Additionally, the authors used the same-sized receptive field derived from two other base networks to capture different characteristics in the extracted features. Deep fusion was further studied in~\cite{zhang2017interleaved,sun2018igcv3,Wang_2020}.

\subsection{Our approach}
 To address the challenges of the existing approaches, we introduce a \ac{DSDF} block that takes two different scale features as input. While propagating information flow in the same resolution, the \ac{DSDF} block also performs a cross resolution fusion. This establishes a dual-scale fusion of features that inherit both high- and low-resolution feature representations. An encoder network is used to feed the feature representations to the \ac{MSRF} sub-network that consists of multiple \ac{DSDF} blocks, thereby performing multi-scale feature exchange. Later, decoder layers with skip-connections from our sub-network and a triple attention mechanism are used to process our fused feature maps together with the shape stream. It is to be noted that the fusion strategy is interchangeable, i.e., low-to-high resolution and vice-versa. 
%
%
\section{The \sysname architecture}
\begin{figure*}[!t]
    \centering
    \includegraphics[width=0.95\textwidth]{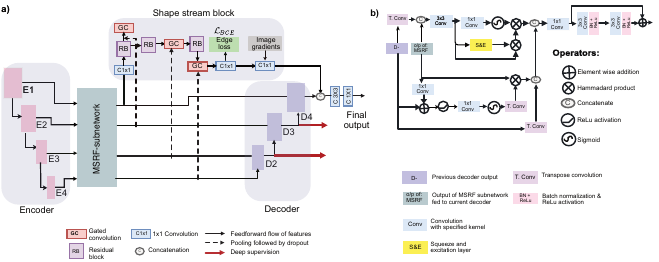}
    \caption{\textbf{The proposed MSRF-Net architectures.} a) Overall block diagram of our network and b) overview of our decoder network.}
    \label{fig:MSRF-Net}
    \vspace{-3mm}
\end{figure*}

Figure~\ref{fig:MSRF-Net}(a) represents the \sysname that consists of an encoder block, the~\ac{MSRF} sub-network, a shape stream block, and a decoder block. The encoder block consists of squeeze and excitation modules, and the \ac{MSRF} sub-network is used to process low-level feature maps extracted at each resolution scale of the encoder. The MSRF sub-network incorporates several \ac{DSDF} blocks. A gated shape stream is applied after the \ac{MSRF} sub-network, and decoders consisting of triple attention blocks are used in the proposed architecture. A triple attention block has the advantage of using spatial and channel-wise attention along with spatially gated attention, where irrelevant features from \ac{MSRF} sub-network are pruned. Below, we briefly describe each component of our \sysname. 
\subsection{Encoder}
The encoder blocks (E1--E4) in Figure~\ref{fig:MSRF-Net}(a) are comprised of two consecutive convolutions followed by a squeeze and excitation module. The $\text{S\&E}$ block in the network increases the network's representative power by computing the interdependencies between channels. During the squeezing step, global average pooling is used to aggregate feature maps across the channel's spatial dimensions. In the excitation step, a collection of per-channel weights are produced to capture channel-wise dependencies~\cite{hu2018squeeze}. At each encoder stage, max pooling with the stride of two is used for downscaling the resolution, and drop out is utilized for the model regularization.  

\subsection{The \ac{DSDF} block and MSRF sub-network}{\label{sec:DSDF}}

Maintaining the resolution throughout the feature encoding process can help the target images become more semantically richer and spatially accurate. The \ac{DSDF} block helps to exchange information between scales, preserve low-level features, and improves information flow while maintaining resolution. The block has two parallel streams for two different resolution scales (Figure~\ref{fig:smfrdbblock}(a)). If we let a $3 \times 3$ convolution followed by a LeakyRelu activation be represented by the operation $\CLR(\cdot)$,  then each stream has a densely connected residual block with five $\CLR$ operations in series. The output feature map $M_{d,h}$ of the $d$-th $\CLR$ operation is computed from the high-resolution input $X_h$ as follows:
\begin{equation}{\label{eq:featuremaphigh}}
    M_{d,h} = \CLR(M_{d-1,h} \oplus M_{d-1,l} \oplus M_{d-2,h}\oplus \cdots \oplus M_{0,h}) 
\end{equation}
\noindent{Here}, $\oplus$ is the concatenation operation, and $h$ represents $\CLR$ operation is on the higher resolution stream of the \ac{DSDF} block. $M_{d-1,l}$ is processed by a transposed convolutional layer with a $3 \times 3$ kernel size and stride of $2$ before being concatenated.
Similarly, for lower resolution stream the output of the $d$-th $\CLR$ operation is denoted by $M_{d,l}$ and represented as:
\begin{equation}{\label{eq:featuremaplow}}
M_{d,l} = \CLR(M_{d-1,l}\oplus M_{d-1,h}\oplus M_{d-2,l}\oplus \cdots \oplus M_{0,l})
\end{equation}
\noindent{Here}, $M_{d-1,h}$ is processed by a convolutional layer with kernel size of $3\times3$ and stride of $2$ before being concatenated.
In Equation~\ref{eq:featuremaphigh} and Equation~\ref{eq:featuremaplow}, $d$ ranges from $1 \leq d \leq 5$. Initially, $X_{h}$ (or $M_{0,h}$) and $X_{l}$ (or $M_{0,l}$) are the higher and lower resolution stream input, respectively. The output of each $\CLR$ has $k$ output channels denoting the growth factor, which regulates the amount of new features the layer can extract and propagate further in the network. Since the growth factor varies for each scale, we only use two scales at once in the \ac{DSDF} to reduce the model's computational complexity for making the training feasible. Further, local residual learning is used to improve information flow, and residual scaling is used to prevent instability~\cite{lim2017enhanced,szegedy2017inception}. Scaling factor $0\leq w \leq 1$ can be used for residual scaling. The final output of the \ac{DSDF} block can be written as (see Figure~\ref{fig:smfrdbblock}(a)): 
\begin{equation}{\label{eq:5}}
X_{r} = w \times {M_{5,r}} + X_{r},
\end{equation}
\noindent{where} $r \in [h, l]$ is the resolution with \textit{h} indicating high-resolution representation and \textit{l} for low resolution representation.

Next, we present an MSRF sub-network that comprises of several \ac{DSDF} blocks to achieve a global multi-scale context using the dual-scale fusion mechanism. As shown in~\cite{zhang2018residual}, our approach has a contiguous memory mechanism that allows retaining multi-scale feature representations since the inputs of each \ac{DSDF} is passed to each subsequent \ac{DSDF} blocks in the same resolution stream.

\begin{algorithm}
\caption{\ac{MSRF} sub-network} 
\begin{algorithmic}[1]
\STATE{Information exchange across all scales in \ac{MSRF} Sub-network}{}\\
\STATE $N$ is no. of DSDF layers ($N=6$ in Figure~\ref{fig:smfrdbblock}(b)) 
\STATE  $H \gets {X_{\hat{h},1}, X_{\hat{h},3},X_{\hat{h}+1,1}, X_{\hat{h}+1,3},...}$ (High-res. input) \\
\STATE  $L \gets {X_{\hat{l},2}, X_{\hat{l},4},X_{\hat{l}+1,2}, X_{\hat{l}+1,4},...}$ (Low-res. input)\\
\STATE  $p\in\{1,3\}$ and $q\in\{2,4\}$ are scale pairs
\STATE $X_{\hat{h}+1,p},X_{\hat{l}+1,q} = \DSDF(X_{\hat{h},p}, X_{\hat{l},q})$
\STATE \textbf{Update:} $\tilde{X}_{\hat{h}, p} = X_{\hat{h}+1,p}$, $\tilde{X}_{\hat{l}, q}$ = $X_{\hat{l}+1,q}$
\FOR {$2\leq $L$ \leq N-3$}
    \STATE $X_{\hat{h}+1,p},X_{\hat{l}+1,q} = \DSDF(X_{\hat{h},p}, X_{\hat{l},q})$
    \STATE\textbf{Update:} $\tilde{X}_{\hat{h}, p},\tilde{X}_{\hat{l}, q} = {X}_{\hat{h}+1,p}, \tilde{X}_{\hat{l}+1,q}$
    \STATE ${X}_{\hat{l}+1,2}, \tilde{X}_{\hat{h}+1,3} = \DSDF(X_{\hat{l},2}, X_{\hat{h},3} )$
    \STATE \textbf{Update:}$\tilde{X}_{\hat{l}, 2},\tilde{X}_{\hat{h}, 3} = {X}_{\hat{l}+1,2}, \tilde{X}_{\hat{h}+1,3}$
\ENDFOR
\STATE $X_{\hat{h}+1,p},X_{\hat{l}+1,q} = \DSDF(X_{\hat{h},p}, X_{\hat{l},q})$
\STATE \textbf{Update:} $\tilde{X}_{\hat{h}, p} = w.X_{\hat{h}+1,p}+X_{0,p}$, $\tilde{X}_{\hat{l}, q}$ = $w.X_{\hat{l}+1,q}+X_{0,q}$
\end{algorithmic}
\label{alg:algo1}
\end{algorithm}

In Algorithm~\ref{alg:algo1}, we define inputs in the MSRF sub-network as the process of demarcating all the resolution scale pairs and feeding them in their respective \ac{DSDF} blocks. For this, we start with the first layer with each layer consisting of four resolution scales with $H$ and $L$ representing a high-resolution and low-resolution set of features, respectively, and each respective block is denoted by $\hat{h}$ and $\hat{l}$. The $\DSDF(\cdot)$ function performs feature fusion across scales in the \ac{DSDF} block, where $\left(X_{\hat{h},p}, X_{\hat{l},q}\right)$ is jointly computed from the $p$ and $q$ scale pairs. Moreover, $\tilde{X}$ represents the feature exchange in the center \ac{DSDF}. 
%
Already after the fourth layer of the MSRF sub-network, we effectively exchange features across all scales and attain global multi-scale fusion (refer to the red rectangular block in Figure~\ref{fig:smfrdbblock}(b)). We can observe that $X_{0,r}, \forall{r}\in \{1,2,3,4\}$ is able to transmit its features to all the parallel resolution representations through multiple \ac{DSDF} blocks. Using this method, we exchange features globally in a more effective way, even when the number of resolution scales is greater than $4$. Similar to the \ac{DSDF} block, the output of the last layer of the sub-network is again scaled by $w$ and added to the original input of the \ac{MSRF} sub-network.
%
\subsection{Shape stream}
We have incorporated the gated shape stream~\cite{takikawa2019gated} in \sysname for the shape prediction (see the shape stream block in Figure~\ref{fig:MSRF-Net}(a)). The \ac{DSDF} blocks can extract relevant high-level feature representations that include important information about shape and boundaries and can be used in the shape stream.  Similar to~\cite{sun2020saunet}, we define $S_{l}$ as the shape stream feature maps where $l$ is the number of layers and $X$ is the output of the \ac{MSRF}-sub-network. Bilinear interpolation is used so that $X$ can match spatial dimensions of $S_{l}$, attention map $\alpha_l$ at the gated convolution is computed as: 
\begin{equation}
\alpha_l = \sigma\left(\conv_{1\times 1}\left(S_{l}\oplus X\right)\right)
\end{equation}
\noindent{where} $\sigma(\cdot)$ is the sigmoid activation function. Finally, $S_{l+1}$ is computed as $S_{l+1} = \text{RB}(S_{l} \times \alpha)$, where $\text{RB}$ represents residual block with two $\CLR$ operations followed by a skip-connection. The output of the shape stream is concatenated with the image gradients of the input image and merged with the original segmentation stream before the last $\CLR$ operation. This is done to increase the spatial accuracy of the segmentation map.
%
\subsection{Decoder}
The decoder block (D2--D4) has skip-connections from the MSRF sub-network and the previous decoder output (say $D^-$) except for D2, where the previous layer connection is the MSRF sub-network output of the E4 (Figure~\ref{fig:MSRF-Net}(a)).
In the decoder block (Figure~\ref{fig:MSRF-Net}(b)), we use two attention mechanisms. The first attention mechanism applies channel and spatial attention, whereas the second attention uses a gating mechanism. We have used a  $\text{S\&E}$ block for the calculation of channel-wise scale coefficients denoted by $X_{\alpha_{se}}$. Spatial attention is also calculated at the same top stream where the input channels $\mathit{C}$ are reduced to $1$ using $1\times1$ convolution. The sigmoid activation function $\sigma(\cdot)$ is used to scale the values between 0 and 1 to produce an activation map, which is stacked $C$ times to give $X_{\alpha_s}$. The output of the spatial and channel attention can be represented as:
\begin{equation}{\label{eq:decodersc}}
D_{sc} = (X_{\alpha_s}+1)\otimes  X_{\alpha_{se}}
\end{equation}
\noindent{where} $\otimes$ denotes the Hadamard product, and $X_{\alpha_s}$ is increased by a magnitude of 1 to amplify relevant features determined by the activation map. We also use the attention gated mechanism~\cite{oktay2018attention}. Let the features coming from MSRF-Net be $X$, and the output from the previous decoder block be $D^-$, then the attention coefficients can be calculated as:
\begin{equation}{\label{eq:dga}}
D_{AG}=\Omega\left(\sigma\left(\Psi(\theta(X)+\phi(D^-))\right)\right)
\end{equation}
%
%
where $\theta(\cdot)$ is the convolution operation with stride 2, kernel size 1, and $G$ channel outputs;
$\phi(\cdot)$ is a convolution operation with stride 1 and kernel size $1\times 1$ applied to $D^-$ giving the same $G$ channels; and
$\Psi(\cdot)$ is convolution function with $1\times 1$ kernel size applied to a combined features from $\theta(\cdot)$ and $\phi(\cdot)$ making output channel equal to 1. Finally, $\sigma(\cdot)$ is applied to obtain the activation map on which transpose convolution operation $\Omega(\cdot)$ is applied. $D_{AG}$ captures the contextual information and identifies the target regions and structures of the image. $\tilde{D}_{AG}= D_{AG} \otimes X$ allows the irrelevant features to be pruned and relevant target structure and regions to be propagated further. $\tilde{D}_{AG}$ is updated as:
\begin{equation}
    \tilde{D}_{AG} = \tilde{D}_{AG} \oplus \Omega(D^-)
\end{equation}
%
%
Now, the final output of the \textit{triple attention decoder block} (i.e., the combination of channel, spatial and gated spatial attention) is $D_\alpha= D_{sc} \oplus  \tilde{D}_{AG}$, which is then followed by two $\CLR$ operations. 

\subsection{Loss computation}
\label{section:loss_comp}
We have used binary cross-entropy loss \lbce  as defined in Equation~\ref{eq:bce} where $y$ is the ground truth value and $\hat{y}$ is the predicted value. We have also used dice loss \ldcs, which is defined in Equation~\ref{eq:dcs}. 

\begin{equation}
    \lbce = (y-1) \log (1 - \hat{y}) - y \log \hat{y} 
   \label{eq:bce}
\end{equation}
\begin{equation}
\ldcs = 1 - \frac{2y\hat{y}+1}{y+\hat{y}+1}
\label{eq:dcs}
\end{equation}

The sum of the two loss functions, $\lcomb = \lambda_{1} \lbce + \lambda_{2} \ldcs$, is used for gradient minimization between the predicted maps and the labels, while only \lbce has been used for shape stream. Here, we set the values of $\lambda_{1}$ and $\lambda_{2}$ to 1. For the latter loss, predicted edge maps and ground truth maps are used during computation. Deep supervision is also used to improve the flow of the gradients and regularization~\cite{lee2015deeply}. Thus, our final loss function can be represented as:
\begin{equation}
\lmsrf = \alpha\lcomb + \beta_{1}\lcomb^{DS^0} +\beta_{2}\lcomb^{DS^1}  + \gamma\lbce^{SS}
\label{eq:losseqn}
\end{equation}
\noindent{where} $\lcomb^{DS^0}$ and  $\lcomb^{DS^1}$ representing the two deep supervision outputs losses (see Figure~\ref{fig:MSRF-Net}(a)) and $\mathcal{L}^{SS}$ is the loss computed for the shape stream. Here, we set the values of $\alpha=1$, $\beta_{1}=1$, $\beta_{2}=1$ and $\gamma=1$ for our experiments.

\section{Experimental setup}
\subsection{Dataset}
To evaluate the effectiveness of the \sysname, we have used four publicly available biomedical imaging datasets; Kvasir-SEG~\cite{jha2020kvasir}, CVC-ClinicDB~\cite{bernal2015wm}, 2018 Data Science Bowl~\cite{caicedo2019nucleus}, and ISIC-2018 Challenge~\cite{codella2018skin,tschandl2018ham10000}. The details about the datasets, number of training and testing samples used, and their availability is presented in Table~\ref{table:datasettable}. All of these datasets consist of the images and their corresponding ground truth masks. An example of each dataset can be found in Figure~\ref{fig:qualitativeresults}. The chosen datasets are commonly used in biomedical image segmentation. The main reason for choosing diverse imaging modalities datasets is to evaluate the performance and robustness of the proposed method.  


\subsection{Evaluation metrics}
Standard computer vision metrics for medical image segmentation such as \acf{DSC}, \ac{mIoU}, recall, precision, and \ac{FPS} have been used for the evaluation of our experimental results. The standard deviations for \ac{DSC}, \ac{mIoU}, r and p are also provided. Additionally, we conduct a paired t-test between the \ac{DSC} achieved by our proposed \sysname and the \ac{DSC} attained by other \ac{SOTA} methods. The p-values of the paired t-tests are also reported.

\subsection{Implementation details}

We have implemented the proposed architecture using the Keras framework~\cite{chollet2015keras} with TensorFlow~\cite{abadi2016tensorflow} as backend. All experiments are conducted on an NVIDIA DGX-2 machine that uses NVIDIA V100 Tensor Core GPUs. The Adam optimizer was used with a learning rate of $0.0001$, and a dropout regularization with $p=0.2$ was used. The scaling factor for our \ac{DSDF} and MSRF sub-network was set to 0.4 ($w=0.4$). The growth factor $k$ is set to 16, 32, and 64 for resolution scale pairs in the \ac{DSDF}. For Kvasir-SEG and 2018~\ac{DSB}, the images are resized to $256\times256$. ISIC-2018 images are resized to  $384\times512$, and images from CVC-ClinicDB are resized to $384\times 288$ resolution. We have used the batch size of 16 for Kvasir-SEG and 2018 DSB, eight for CVC-ClinicDB, and four for the ISIC-2018 Challenge dataset. We have empirically set the number of epochs for all datasets to 200 epochs. We have used 80\% of the dataset for training, 10\% for validation, and the remaining 10\% for testing. Data augmentation techniques such as random cropping, random rotation, horizontal flipping, vertical flipping, and grid distortion were applied. It is to be noted that we have used open-source code provided by the respective authors for all the baseline comparisons. The proposed model is available at \href{https://github.com/NoviceMAn-prog/MSRF-Net}{https://github.com/NoviceMAn-prog/MSRF-Net}. 
\vspace{-3mm}

\section{Results}

\subsection{SOTA method comparisons}
In this section, we present the comparison of our \sysname with other \ac{SOTA} methods.

\begin{table} [t!]
 \caption{{The medical datasets used in our experiments. }}
    \label{table:datasettable}
   \footnotesize
    \centering
    \begin{tabular}{p{2.8cm}|p{0.8cm}|p{1.2cm}|p{0.5cm}|p{0.5cm}|p{0.5cm}}
            \toprule
        \textbf{Dataset} & \textbf{Images} & \textbf{Input size} & \textbf{Train} & \textbf{Valid}& \textbf{Test}\\ 
              \hline
              \hline
        Kvasir-SEG~\cite{jha2020kvasir}  & 1000 & Variable &800 &100 &100\\  \hline
        CVC-ClinicDB~\cite{bernal2015wm} & 612 & $384\times 288$ &490  & 61& 61 \\  \hline
        2018 Data Science Bowl~\cite{caicedo2019nucleus} & 670 & $256\times 256$ &536 &67 & 67\\ \hline
        ISIC-2018 Challenge~\cite{codella2018skin,tschandl2018ham10000}& 2596 & $384 \times 512$ &2078 &259 &259 \\ 
        \hline
        \bottomrule
\end{tabular}
\vspace{-5mm}
\end{table}	
\subsubsection{Comparison on Kvasir-SEG}

\begin{table*}[!t]
\centering
\footnotesize
\caption{Result comparison on the Kvasir-SEG dataset. We have not computed paired t-test (p values) for the same network (MSRF-Net).}
\begin{tabular}{@{}l|l|l|l|l|l|l|l@{}}
\toprule
\textbf{Method} & \textbf{DSC} & \textbf{mIoU} & \textbf{Recall} & \textbf{Precision} & \textbf{P-values} & \textbf{Parameters} & \textbf{FPS}\\ 
\hline
\hline
U-Net~\cite{ronneberger2015u} & 0.8629 ± 0.2334 & 0.8176 ± 0.2465 & 0.9094 ± 0.2216 &0.8901 ± 0.2313 & 1.559e-02  & 7.11M & 41.04  \\ \hline
U-Net++~\cite{zhou2019unet++} & 0.7475 ± 0.2664 & 0.6313 ± 0.2788 & 0.6865 ± 0.2888 & 0.8871 ± 0.2689 & 4.363e-10 & 9.04M & 30.67 \\ \hline
ResUNet++~\cite{jha2019resunet++} & 0.8189 ± 0.2652 & 0.7918 ± 0.2819 & 0.8372 ± 0.2751 & 0.9255 ± 0.2545 & 6.518e-06 & 4.07M & 15.92\\ \hline

Deeplabv3+( Xception)~\cite{chen2018encoder} & 0.8965 ± 0.2072 & 0.8575 ± 0.2290  & 0.8984 ± 0.2099  & 0.9496 ± 0.1801 & 3.660e-01 & 41.25M & 49.11 \\ \hline

Deeplabv3+ (Mobilenet)~\cite{chen2018encoder} & 0.8656 ± 0.2032 & 0.8186 ± 0.2222 & 0.8808 ± 0.2105 & 0.9205 ± 0.1980 & 1.409e-04 & 2.14M & 118.50\\ \hline

{DoubleUNet}~\cite{jha2020doubleu} &0.8699 ± 0.1585 & {0.8166} ± {0.1933} &{0.9039} ± 0.1810 & 0.8942 ± 0.1586 & 8.109e-05 & 29.29M & 7.46 \\ \hline

HRNetV2-W18-Smallv2~\cite{Wang_2020} & 0.8179 ± 0.2067 & 0.7470 ± 0.2622  & 0.8016 ± 0.2681 & 0.8696 ± 0.2494 & 4.844e-06 & 26.20M & 52.68 \\ \hline

HRNetV2-W48~\cite{Wang_2020} & 0.8896 ± 0.1200 & 0.8262 ± 0.1856 & 0.8973 ± 0.1719 & 0.9056 ± 0.1492 & 5.935e-04 & 65.84M & 29.79 \\ \hline

ColonSegNet~\cite{jha2021real} & 0.8203 ± 0.2295 & 0.7435 ± 0.2539 & 0.8124 ± 0.2494  & 0.8832 ± 0.1985 & 8.692e-07 & 5.01M & 24.26 \\ \hline

DDANet~\cite{tomar2020ddanet}& 0.8915 ± 0.1880 & 0.8393 ± 0.2126 & 0.8927 ± 0.2093 & 0.9213 ± 0.1604 & 2.558e-02 & 6.83M & 7.76\\ \hline

ResUNet++ + CRF$^\diamond$~\cite{jha2021comprehensive}&0.7965 ± 0.2707 & 0.8250 ±0.2832 &0.8119 ± 0.2803  &0.8045 ± 0.2638 & 1.721e-06 & 4.02M & 15.12 \\ \hline

PraNet~\cite{fan2020pranet}  & 0.9078 ± 0.1543 & 0.8561 ± 0.1823 & 0.9034 ± 0.1719 & 0.9352 ± 0.1358 & 3.693e-01 & 32.54M & 48.25  \\ \hline

UACANet-S~\cite{kim2021uacanet} & 0.8800 ± 0.2042 & 0.8250 ± 0.2215 & 0.8701 ± 0.2136 & 0.9229 ± 0.1758 & 3.871e-04 & 26.90M & 32.58 \\ \hline

UACANet-L~\cite{kim2021uacanet} & 0.9014 ± 0.1878 & 0.8555 ± 0.2098 & 0.8897 ± 0.2148 & 0.9381 ± 0.1458 & 4.878e-01 & 69.15M & 28.40 \\ \hline

\sysname(Ours) & \textbf{0.9217} ± 0.1685 & \textbf{0.8914} ± 0.1938 & \textbf{0.9198} ± 0.1919 & \textbf{0.9666} ± 0.1379 & - & 18.38M & 14.38\\ \hline
\bottomrule
\end{tabular}
\label{tab:result1}
\vspace{-5mm}
\end{table*}

Early detection of polyps, before they potentially change into colorectal cancer, can improve the survival rate~\cite{levin2008screening}. Therefore, we have selected two popular colonoscopy datasets in our experiment. The first colonoscopy dataset is Kvasir-SEG. We report the quantitative evaluation of \sysname in Table~\ref{tab:result1} and qualitative results in Figure~\ref{fig:qualitativeresults}. From the quantitative results, we can observe that our method outperforms all the other \ac{SOTA} methods on all metrics. It achieves 1.39\% improvement on \ac{DSC} as compared to PraNet~\cite{fan2020pranet}, 3.39\% improvement on \ac{mIoU} as compared Deeplabv3+ with Xception backbone~\cite{chen2018encoder}. Our method also achieves an improvement of 1.70\% on precision and 1.04\% on recall as compared to Deeplabv3+ with Xception backbone and U-Net~\cite{ronneberger2015u}, respectively. The network's ability to segment polyps can be observed from the ground truth comparison with the predicted mask. (Figure~\ref{fig:qualitativeresults}). 

\subsubsection{Comparison on CVC-ClinicDB}
\begin{table*}[!t]
\centering
\footnotesize
\caption{Result comparison on the CVC-ClinicDB.}
\label{tab:result2}
\begin{tabular}{@{}l|l|l|l|l|l|l|l@{}}
\toprule
\textbf{Method} &\textbf{DSC}   &\textbf{mIoU} &\textbf{Recall} &\textbf{Precision}& \textbf{P-values} & \textbf{Parameters} & \textbf{FPS} \\
\hline
\hline

U-Net~\cite{ronneberger2015u}  & 0.9145 ± 0.1390 & 0.8654 ± 0.1514 &0.9178 ± 0.1309 &0.9381 ± 0.1594 & 1.000e+00$^\star$ & 7.11M & 22.84    \\ \hline

U-Net++~\cite{zhou2019unet++} & 0.8453 ± 0.1516 & 0.7559 ± 0.1188 & 0.8917 ± 0.2594 & 0.8323 ± 0.2713 & 6.400e-04 & 9.04M & 17.60 \\ \hline

ResUNet++$^\diamond$~\cite{jha2019resunet++} & 0.9075 ± 0.1455 & 0.8587 ± 0.1616 & 0.9156 ± 0.1372 & 0.9325 ± 0.1593 & 1.000e+00$^\star$ & 4.07M & 15.71 \\ \hline

Deeplabv3+ (Xception)~\cite{chen2018encoder} & 0.8897 ± 0.1895 & 0.8706 ± 0.2036 & 0.9251 ± 0.1965 & 0.9366 ± 0.1621 & 6.723e-01 & 41.25M & 29.08 \\\hline

Deeplabv3+ (Mobilenet)~\cite{chen2018encoder} & 0.8985 ± 0.1385 & 0.8588 ± 0.1544 & 0.9160 ± 0.1562 & 0.9287 ± 0.1378 & 2.838e-01 & 2.14M & 55.68 \\ \hline

DoubleU-Net~\cite{jha2020doubleu} & 0.9272 ± 0.1761 & 0.8889 ± 0.1896 & 0.9395 ± 0.1800 & \textbf{0.9592} ± 0.1609 & 1.000e+00$^\star$ & 29.29M & 7.46 \\ \hline

HRNetV2-W18-Smallv2~\cite{Wang_2020} & 0.9073 ± 0.1158 & 0.8457 ± 0.1477 & 0.9137 ± 0.1293 & 0.9191 ± 0.1060 & 1.000e+00$^\star$ & 26.20M & 57.47 \\ \hline 

HRNetV2-W48~\cite{Wang_2020} & 0.9244 ± 0.1280 & 0.8747 ± 0.1332 & 0.9234 ± 0.1356 & 0.9296 ± 0.1318 & 1.000e+00$^\star$ & 65.84M & 29.76 \\ \hline

ColonSegNet~\cite{jha2021real} & 0.9132 ± 0.1416 & 0.8600 ± 0.1585 & 0.9072 ± 0.1564  & 0.9292 ± 0.1393 & 1.000e+00$^\star$ & 5.01M & 7.98 \\ \hline

DDANet~\cite{tomar2020ddanet} & 0.9233 ± 0.1339 & 0.8747 ± 0.1438 & 0.9271 ± 0.1285 & 0.9259 ± 0.1495 & 1.000e+00$^\star$ & 6.83M & 58.15 \\ \hline
ResUNet++ + CRF$^\diamond$~\cite{jha2021comprehensive}&0.8815 ± 0.1507 &0.8899 ± 0.1685 &0.8970 ± 0.1581  &0.8674 ± 0.1605 & 1.102e-01 & 4.02M & 8.55 \\ \hline

PraNet~\cite{fan2020pranet}  & 0.9072 ± 0.1695 & 0.8575 ± 0.1823 & 0.9227 ± 0.1657 & 0.9134 ± 0.1614 & 1.000e+00$^\star$ & 32.54M & 47.92 \\ \hline

UACANet-S~\cite{kim2021uacanet} & 0.9190 ± 0.1426 & 0.8700 ± 0.1556 & 0.9285 ± 0.1356 & 0.9201 ± 0.1574 & 1.000e+00$^\star$ & 26.90M & 31.79 \\ \hline

UACANet-L~\cite{kim2021uacanet}& 0.9098 ± 0.1829 & 0.8649 ± 0.1872 & 0.9174 ± 0.1772 & 0.9114 ± 0.1950 & 1.000e+00$^\star$ & 69.15M & 32.81 \\ \hline

\sysname (Ours) & \textbf{0.9420} ± 0.0804 & \textbf{0.9043} ± 0.1009 & \textbf{0.9567} ± 0.0620 & 0.9427 ± 0.0994 & - & 18.38M & 12.50  \\
\hline
\bottomrule
 \multicolumn{7}{l}{$^\star$ \textit{large improvements were still observed for some samples with consistent Dice scores (also see standard deviation) for \sysname} \hspace{.1cm}}
\end{tabular}
\vspace{-1mm}
\end{table*}

\begin{figure*}[t!]
\centering
    \includegraphics[height=0.38\textwidth]{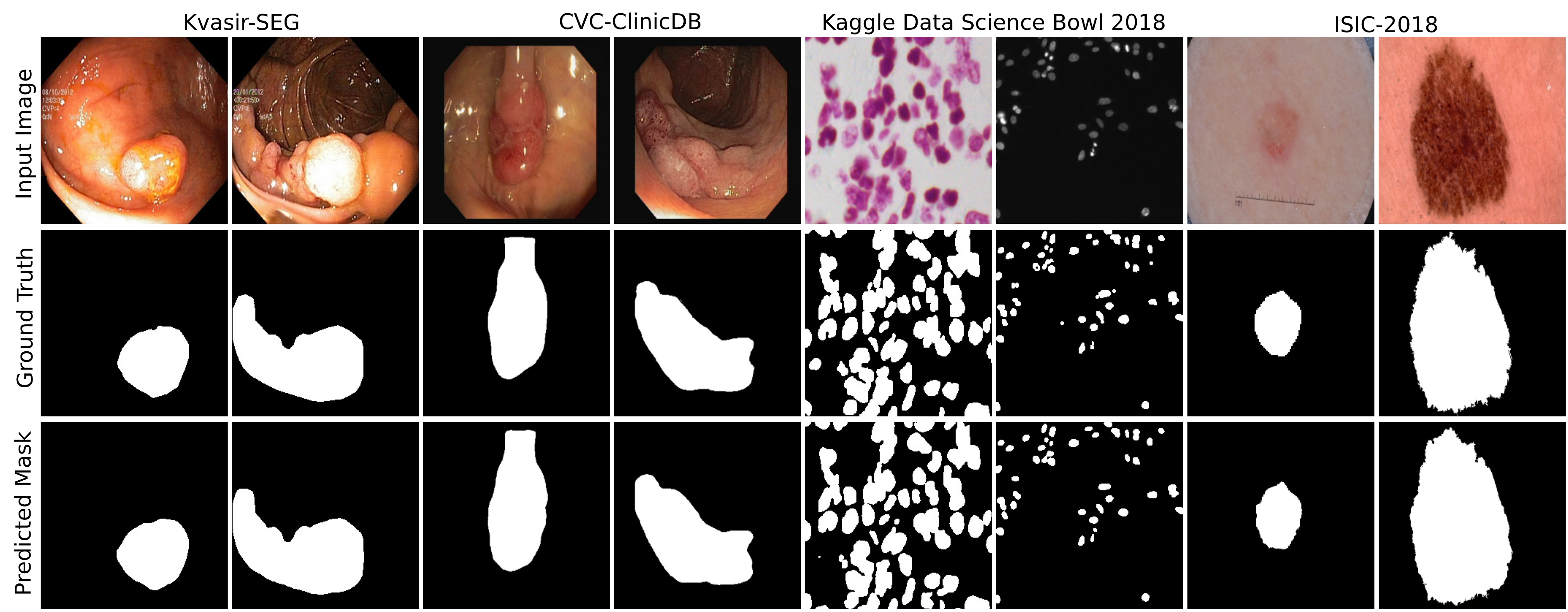}
   \caption{The Figure shows qualitative results of the \sysname on four biomedical imaging datasets.
    \label{fig:qualitativeresults}}
    \vspace{-2mm}
\end{figure*}

\begin{figure}[t!]
    \begin{center}
    
    \includegraphics[width=0.45\textwidth]{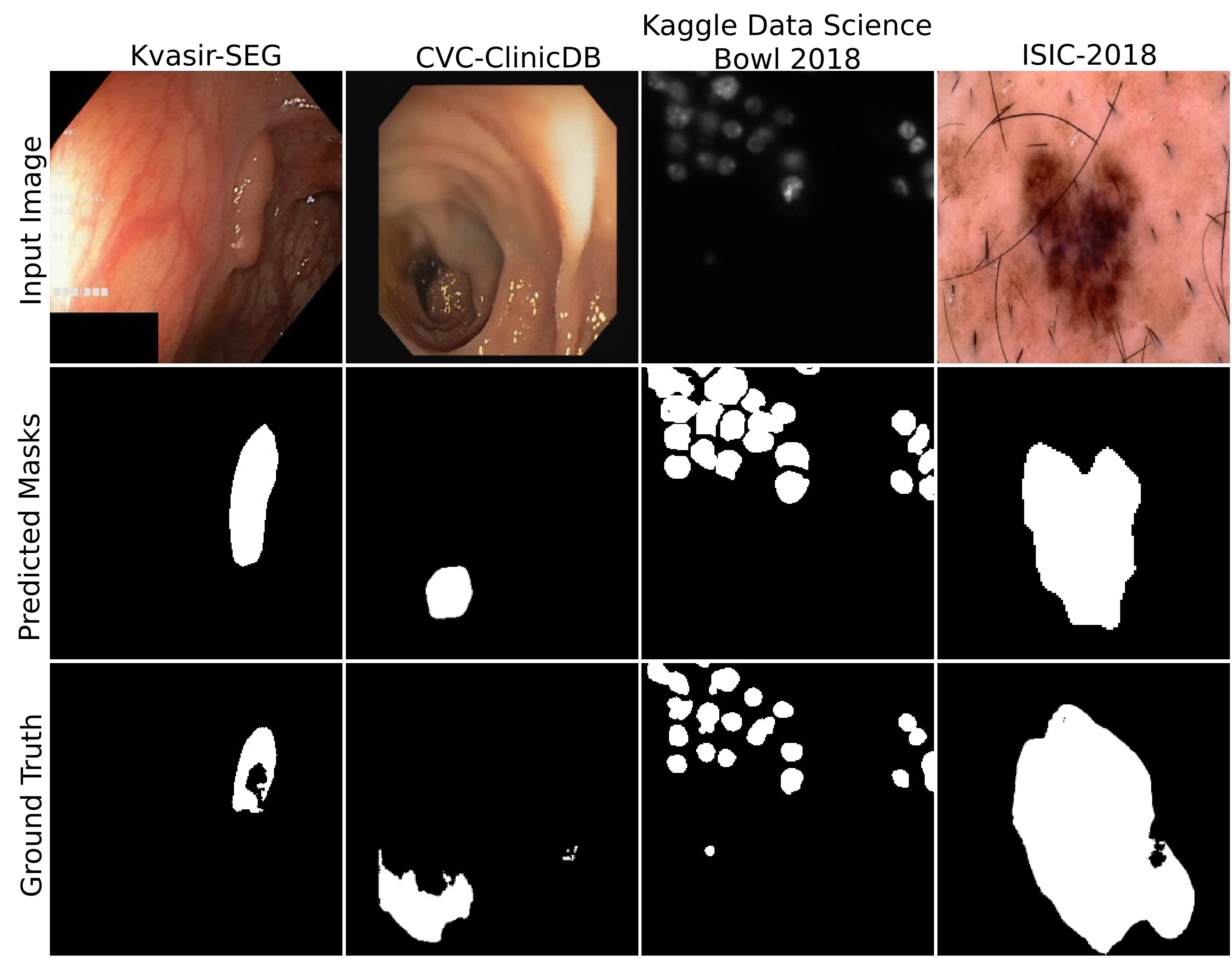}
    \caption{Qualitative results of the \sysname for sub-optimal cases.}
    \label{fig:suboptimal}
    \end{center}
    \vspace{-5mm}
\end{figure}

CVC-ClinicDB is the second colonoscopy dataset used in our experiment. The quantitative results from Table~\ref{tab:result2} show that our approach surpasses other \ac{SOTA} methods and achieves a \ac{DSC} of 0.9420 ± 0.0804, which is 1.76\% improvement in \ac{DSC} over the best-performing HRNetV2-W48~\cite{Wang_2020}. We report a \ac{mIoU} of 0.9043  ± 0.1009 and a recall of 0.9567 ± 0.0620, which is 1.44\% improvement in \ac{mIoU} and 2.82\% improvement in recall over \ac{SOTA} combination of ResUNet++ and conditional random field~\cite{jha2021comprehensive} and UACANet-S~\cite{kim2021uacanet}, respectively. Additionally, \sysname achieves a precision of 0.9427, which is competitive with the best performing DoubleUNet~\cite{jha2020doubleu}. Our method produces prediction masks with nearly the same boundaries and shape of the polyp as compared to the ground truth masks (Figure~\ref{fig:qualitativeresults}). 

\subsubsection{Comparison on 2018 Data Science Bowl}
Finding nuclei in a cell from a large variety of microscopy images is a challenging problem. We experiment with the 2018 Data Science Bowl challenge dataset. Table~\ref{tab:result3} shows the comparison of the result of the proposed \sysname with some of the presented approaches. \sysname obtains a \ac{DSC} of 0.9224 ± 0.0538, \ac{mIoU} of 0.8534 ± 0.0870, recall of 0.9402 ± 0.0734 and precision of 0.9022 ± 0.0601 which outperforms the best performing ColonSegNet~\cite{jha2021real} in most metrics (see Table~\ref{tab:result3}). From the qualitative results in Figure~\ref{fig:qualitativeresults}, we observe that the predicted masks are visually similar to the ground truth masks. 
%
\subsubsection{Comparison on ISIC-2018 Skin Lesion Segmentation challenge}
An automatic diagnosis tool for skin lesions can help in accurate melanoma detection, which is also a commonly occurring cancer and can save life up to 99\%~\cite{cancerfigures} of cases. The quantitative results for the ISIC-2018 challenge are shown in Table~\ref{tab:result4}. Our method achieved a \ac{DSC} of 0.8824 ± 0.1602, \ac{mIoU} of 0.8373 ± 0.1818, recall of 0.8893 ± 0.1889, and precision of 0.9348 ± 0.1488. We can observe an improvement of 0.43\% and 1.37\% over Deeplabv3+ with the Mobilenet backbone~\cite{chen2017deeplab} in \ac{DSC} and \ac{mIoU}, respectively. We also observe a 0.63\% improvement in recall over Deeplabv3+(MobileNet)~\cite{chen2017deeplab}. Our results are comparable with DoubleU-Net~\cite{jha2020doubleu} which reports the highest \ac{DSC} of 0.8938. A higher recall shows that our method is more medically relevant, which is considered as the major strength of our architecture~\cite{gilvary2019missing}. From Figure~\ref{fig:qualitativeresults}, we can observe that our method can segment skin lesions of varying sizes accurately.
\vspace{-0.3cm}
\subsection{Generalization study}\label{section:general}
To ensure generalizability, we have trained our model and other \ac{SOTA} methods on one dataset and then experimented on a new unseen dataset which comes from a different institution, consisting of different cohort populations and acquired using different imaging protocols. To this end, we have used the Kvasir-SEG collected in Vestre Viken Health Trust in Norway for training and tested our trained model on the CVC-ClinicDB, which was captured in Hospital Clinic in Barcelona, Spain. Similarly, we conducted this study on an opposite set-up as well, i.e., training on CVC-ClinicDB and testing on Kvasir-SEG. 

\subsubsection{{Generalizability} results on CVC-ClinicDB}
Table~\ref{tab:generalizationkvasir} shows the generalizability results of the \sysname model trained on Kvasir-SEG and tested on CVC-ClinicDB. Despite using two datasets acquired using two different imaging protocols, \sysname obtained an acceptable \ac{DSC} of 0.7921 ± 0.2564, \ac{mIoU} of 0.6498 ± 0.2729, recall of 0.9001 ± 0.2980, and precision of 0.7000 ± 0.1572. We observe that our \sysname performs better than other \ac{SOTA} methods in terms of \ac{DSC}. HRNetV2-W48~\cite{Wang_2020} obtained a competitive \ac{DSC} of 0.7901.

\subsubsection{{Generalizability} results on  Kvasir-SEG}
Similarly, we present the results of the models trained on CVC-ClinicDB and tested on Kvasir-SEG in \mbox{Table~\ref{tab:generalizationcvc}}. We report that our model achieves a \ac{DSC} of 0.7575 ± 0.2643, \ac{mIoU} of 0.6337 ± 0.2815, recall of 0.7197 ± 0.2775 and precision of 0.8414 ± 0.2731, which outperforms other \ac{SOTA} methods in \ac{DSC} and \ac{mIoU}. The second best performing method is PraNet~\cite{fan2020pranet} with \ac{DSC} of 0.7293 ± 0.3004, and \ac{mIoU} of 0.6262 ± 0.3128. Our method outperforms PraNet~\cite{fan2020pranet} by 2.82\% in \ac{DSC} and 0.75\% in \ac{mIoU} but PraNet~\cite{fan2020pranet} records the highest recall of 0.8007.
\vspace{-0.3cm}
\subsection{Ablation study}{\label{sec:ablationStudy}}
We have conducted an extensive ablation study on the Kvasir-SEG. For this, we ablated the impact of the MSRF sub-network, scaling mechanism used in the network,  the effect of the number of DSDF blocks used, the impact of the MSRF sub-network on shape prediction in the shape stream in Section~\ref{sec:ablationStudy}, the effect observed when shape stream and deep supervision is removed from the architecture of \sysname.

Table~\ref{table:ablationstudy} shows the quantitative results of our ablation study. Initially, we removed the MSRF sub-network which resulted in the least \ac{DSC} of 0.8771. The addition of a subset of the original MSRF sub-network (The red dotted region in Figure~\ref{fig:smfrdbblock}(b)) raises the \ac{DSC} to 0.8986. Further, we removed the \ac{DSDF} with second and third scale inputs (also see Figure~\ref{fig:smfrdbblock}(b), where middle \ac{DSDF} blocks represent them, i.e., layer three and layer five are removed) from the original MSRF sub-network to achieve a \ac{DSC} of 0.9013. To further investigate the contribution of our MSRF sub-network, we remove the shape stream to achieve a \ac{DSC} of 0.9194 which is comparable to highest 0.9217 \ac{DSC} reported by the original \sysname configuration. We disable the triple attention mechanism in the decoder block to get a \ac{DSC} 0.9067 ± 0.1834. Our ablation on further removing deep supervision resulted in a lower \ac{DSC} of 0.8988.
We also report the effect of using a combination of dice loss and binary cross entropy loss in \lcomb used in Equation~\ref{eq:losseqn} to supervise \sysname during training. First, we set  $\lcomb = \lbce$ and secure a \ac{DSC} of 0.9059 which was followed by setting $\lcomb = \ldcs$ which scored a \ac{DSC} of 0.8861. A similar trend was observed for other metrics.

\begin{table*}[!t]
\centering

\scriptsize
\caption{Results on the 2018 Data Science Bowl}
\label{tab:result3}
\begin{tabular}{@{}l|l|l|l|l|l|l|l@{}}
\toprule
\textbf{Method} & \textbf{DSC} & \textbf{mIoU}& \textbf{Recall} & \textbf{Precision} & \textbf{P-values} & \textbf{Parameters} & \textbf{FPS}\\ 
\hline
\hline
U-Net~\cite{ronneberger2015u} & 0.9080 ± 0.0638 & 0.8314 ± 0.1019 & 0.9029 ± 0.0981 & 0.9130 ± 0.0719 & 1.308e-03 & 7.11M & 25.02    \\ \hline
U-Net++~\cite{zhou2019unet++}  & 0.7705 ± 0.3010 & 0.5265 ± 0.3078 & 0.7159 ± 0.3171& 0.6657 ± 0.2745 & 4.832e-04 & 9.04M & 21.86\\ \hline
ResUNet++~\cite{jha2019resunet++} & 0.9098 ± 0.0797 & 0.8370 ± 0.1154 & 0.9169 ± 0.0947 & 0.9057 ± 0.0853 & 2.420e-01 & 4.07M & 19.81 \\ \hline
Deeplabv3+ (Xception)~\cite{chen2018encoder} & 0.8857 ± 0.1674 & 0.8367 ± 0.1702 & 0.9141 ± 0.1751 & 0.9081 ± 0.1689 & 4.641e-01 & 41.25M & 16.20 \\\hline
Deeplabv3+ (Mobilenet)~\cite{chen2018encoder} & 0.8239 ± 0.1613 & 0.7402 ± 0.1618 & 0.8896 ± 0.1720 & 0.8151 ±  0.1657 & 1.044e-06 & 2.14M & 23.70 \\ \hline
DoubleUNet~\cite{jha2020doubleu} & 0.9109 ± 0.0876 & 0.8429 ± 0.1109 &  0.9278 ± 0.0962 & 0.9020 ± 0.0997 & 8.929e-02 & 29.29M & 7.47 \\ \hline
HRNetV2-W18-Smallv2~\cite{Wang_2020} & 0.8495 ± 0.3267 & 0.7585 ± 0.1521 & 0.8640 ± 0.1659 & 0.8398 ± 0.1602 & 4.978e-04 & 26.20M & 58.03  \\ \hline 
HRNetV2-W48~\cite{Wang_2020} & 0.8488 ± 0.1470 & 0.7588 ± 0.1499 & 0.8359 ± 0.1618 & 0.8913 ± 0.0550 & 8.460e-05 & 65.84M & 29.41  \\ \hline
ColonSegNet~\cite{jha2021real} & 0.9197 ± 0.0605 & 0.8466 ± 0.0953 & 0.9153 ± 0.0917  & \textbf{0.9312} ± 0.0532 & 6.433e-01 & 5.01M & 16.56  \\ \hline
DDANet~\cite{tomar2020ddanet} & 0.9182 ± 0.0684 & 0.8452 ± 0.1037  & 0.9139 ± 0.0964 & 0.9289 ± 0.0575 & 5.922e-01 & 6.83M & 19.02 \\ \hline
ResUNet++ + CRF~\cite{jha2021comprehensive}& 0.7806 ± 0.2223 & 0.7322 ± 0.2386 & 0.7534 ± 0.2558 & 0.6308 ± 0.1752 & 6.971e-06 & 4.02M & 72.78 \\ \hline
PraNet~\cite{fan2020pranet}  &0.8751 ± 0.0871 & 0.7868 ± 0.1169 & 0.9182 ± 0.0736 & 0.8438 ± 0.1138 & 1.553e-07 & 32.54M & 11.88 \\ \hline
UACANet-S~\cite{kim2021uacanet} &0.8687 ± 0.0913 & 0.7774 ± 0.1208 & 0.9092 ± 0.0960& 0.8385 ± 0.1115 & 2.940e-08 & 26.90M & 28.08 \\ \hline
UACANet-L~\cite{kim2021uacanet}&0.8688 ± 0.0999 &0.7791 ± 0.1283 &0.9061 ± 0.1057 & 0.8414 ± 0.1141 & 8.975e-07 & 69.15M & 32.26 \\ \hline
\sysname (Ours)  &\textbf{0.9224} ± 0.0538 &\textbf{0.8534} ± 0.0870  & \textbf{0.9402} ± 0.0734  & 0.9022 ± 0.0601 & - & 18.38M & 6.84 \\ \hline 
\bottomrule
\end{tabular}
\vspace{-5mm}
\end{table*}

\begin{table*}[!t]
\centering
\scriptsize
\caption{Results on the ISIC-2018 skin lesion segmentation challenge}
\label{tab:result4}
\begin{tabular}{@{}l|l|l|l|l|l|l|l@{}}
\toprule
\textbf{Method} & \textbf{DSC}  & \textbf{mIoU}& \textbf{Recall} & \textbf{Precision}& \textbf{P-values} & \textbf{Parameters} & \textbf{FPS}\\  \hline
\hline
\hline
U-Net~\cite{ronneberger2015u} & 0.8554 ± 0.1848  & 0.7847 ± 0.2094 & 0.8204 ± 0.2186 & \textbf{0.9474} ± 0.1296 & 1.315e-02 & 7.11M & 79.43 \\ \hline
U-Net++~\cite{zhou2019unet++}  & 0.8094 ± 0.2261 & 0.7288 ± 0.2452 & 0.7866 ±  0.2369 & 0.9084 ± 0.2222 & 6.336e-08 & 9.04M & 60.44 \\ \hline
ResUNet++~\cite{jha2019resunet++} & 0.8557 ± 0.2014 & 0.8135 ± 0.2210 & 0.8801 ± 0.2320 & 0.8676 ± 0.1562 & 1.813e-02 & 4.07M & 40.93 \\ \hline
Deeplabv3+ (Xception)~\cite{chen2018encoder} & 0.8772 ± 0.1465 & 0.8128 ± 0.1806 & 0.8681 ± 0.1792 & 0.9272 ± 0.13602 & 4.314e-02 & 41.25M & 43.53 \\\hline
Deeplabv3+ (Mobilenet)~\cite{chen2018encoder} & 0.8781 ± 0.1371 & 0.8236 ± 0.1711 & 0.8830 ± 0.1725 & 0.9244 ± 0.1317 & 9.503e-02 & 2.14M & 61.10 \\ \hline
HRNetV2-W18-Smallv2~\cite{Wang_2020} & 0.8561 ± 0.3696 & 0.7821 ± 0.2091 & 0.8556 ± 0.2029 & 0.8974 ± 0.1862 & 8.780e-03 & 26.20M & 57.47 \\ \hline 
HRNetV2-W48~\cite{Wang_2020} & 0.8667 ± 0.2453 & 0.8109 ± 0.2630 & 0.8584 ± 0.2936 & 0.9155 ± 0.2755 & 4.270e-02 & 65.84M & 28.54 \\ \hline

DoubleU-Net~\cite{jha2020doubleu} &{\textbf{0.8938}} ± 0.1362 & {0.8212} ± {0.1659} & 0.8780 ± {0.1573} & {0.9459} ± {0.1353} & 1.000e+00 & 29.29M & {7.46}  \\ \hline

ResUNet++ + CRF~\cite{jha2021comprehensive}& 0.8688 ± 0.1719 & 0.8209 ± 0.1971 & 0.8826 ± 0.2063 & 0.8736 ± 0.1540 & 1.813e-02 & 4.02M & 79.11 \\ \hline
MSRF-Net (Ours) & 0.8824 ± 0.1602 & \textbf{0.8373} ± 0.1818 &\textbf{0.8893} ± 0.1889 & 0.9348 ± 0.1488 & - & 18.38M & 16.10 \\ \hline   
\bottomrule
\end{tabular}
\vspace{-5mm}
\end{table*}

\begin{table*} [!t]
\centering
\scriptsize
\caption{Generalizability results of the models trained on Kvasir-SEG and tested on  CVC-ClinicDB}
\label{tab:generalizationkvasir}
\begin{tabular}{l|r|r|r|r|r|r}
\toprule
\textbf{Method} & \textbf{DSC} & \textbf{mIoU}& \textbf{Recall} & \textbf{Precision} & \textbf{Parameters} & \textbf{FPS} \\\hline
U-Net~\cite{ronneberger2015u} & 0.7172 ± 0.2911 & 0.6133 ± 0.2870 & 0.7255 ± 0.3246 & \textbf{0.7986} ± 0.2775 & 7.11M & 24.15\\ \hline

U-Net++~\cite{zhou2019unet++}  & 0.4265 ± 0.3922  & 0.3345 ± 0.3518 & 0.3939 ± 0.4480 & 0.6894 ± 0.4111   & 9.04M & 21.18     \\ \hline
ResUNet++~\cite{jha2019resunet++} & 0.5560 ± 0.3436 & 0.4542 ± 0.3174 & 0.5795 ± 0.3896 & 0.6775 ± 0.3579 & 4.07M & 8.84 \\ \hline

Deeplabv3+ (Xception)~\cite{chen2018encoder} & 0.6509 ± 0.3172 & 0.5385 ± 0.3174 & 0.6251 ± 0.3621 & 0.7947 ± 0.3175 & 41.25M &  27.24\\\hline

Deeplabv3+ (Mobilenet)~\cite{chen2018encoder} & 0.6303 ± 0.2740 & 0.4825 ± 0.2716 & 0.5957 ± 0.3391 & 0.7173 ± 0.2730 & 2.10M & 73.64\\ \hline

HRNetV2-W18-Smallv2~\cite{Wang_2020} & 0.6428 ± 0.3003 & 0.5513 ± 0.3213 & 0.6811 ± 0.3753 & 0.7253 ± 0.3191 & 26.20M & 57.32 \\ \hline 

HRNetV2-W48~\cite{Wang_2020} & 0.7901 ± 0.2280 & \textbf{0.6953} ± 0.2455  & 0.8796 ± 0.1746 & 0.7694 ± 0.2642 & 65.84M & 29.24\\ \hline

ColonSegNet~\cite{jha2021real} & 0.6895 ± 0.2716 & 0.5813 ± 0.2754 & 0.7862 ± 0.2965  & 0.7177 ± 0.2897 & 5.01M & 17.01\\ \hline

ResUNet++ + CRF~\cite{jha2021comprehensive}& 0.6502 ± 0.3381 & 0.7417 ± 0.3147 & 0.7047 ± 0.3631 & 0.6277 ± 0.3392 & 4.02M & 80.31  \\ \hline

PraNet~\cite{fan2020pranet}  & 0.7225 ± 0.2931 & 0.6328 ± 0.3028 & 0.7531 ± 0.3390 & 0.7888 ± 0.2953 & 32.54M & 30.91  \\ \hline

UACANet-S~\cite{kim2021uacanet} & 0.5683 ± 0.3799 & 0.4907 ± 0.3631 & 0.5792 ± 0.4101 & 0.7095 ± 0.3833 & 26.90M & 32.61 \\ \hline

UACANet-L~\cite{kim2021uacanet}& 0.5589 ± 0.3899  & 0.4849 ± 0.3689 & 0.5800 ± 0.4276 & 0.6775 ± 0.3906 & 69.15M & 32.20 \\ \hline

MSRF-Net (Ours) & \textbf{0.7921} ± 0.2564 & 0.6498 ± 0.2729 & \textbf{0.9001} ± 0.2980 & 0.7000 ± 0.1572 & 18.38M & 10.94\\ \hline   
\bottomrule
\end{tabular}
\vspace{-5mm}
\end{table*}
\begin{table*}[!t]
\centering
\scriptsize
\caption{Generalizability results of the models trained on CVC-ClinicDB and tested on Kvasir-SEG}
\label{tab:generalizationcvc}
\begin{tabular}{l|r|r|r|r|r|r}
\toprule
\textbf{Method} & \textbf{DSC} & \textbf{mIoU}& \textbf{Recall} & \textbf{Precision}& \textbf{Parameters} & \textbf{FPS}  \\ 
\hline
\hline
U-Net~\cite{ronneberger2015u} & 0.6222 ± 0.2595 & 0.4588 ± 0.2609 & 0.5129 ± 0.1766 & 0.8133 ± 0.3059 & 7.11M & 35.18\\ \hline

U-Net++~\cite{zhou2019unet++}  & 0.5926 ± 0.2363  & 0.4564 ± 0.2321 & 0.7352 ± 0.2368 & 0.5462 ± 0.2902  & 9.04M & 16.85      \\ \hline

ResUNet++~\cite{jha2019resunet++} & 0.5147 ± 0.3138 & 0.4082 ± 0.2940 & 0.7181 ± 0.3331 & 0.4860 ± 0.3530 & 4.07M & 25.86 \\ \hline

HRNetV2-W18-Smallv2~\cite{Wang_2020} & 0.7012 ± 0.0680 & 0.6009 ± 0.2889 & 0.7184 ± 0.3021 & 0.7666 ± 0.2863  & 26.20M & {57.89} \\ \hline 

HRNetV2-W48~\cite{Wang_2020} & 0.7404 ± 0.1489 & 0.6233 ± 0.2834 & 0.7293 ± 0.2830 & \textbf{0.8511} ± 0.2563 & 65.84M & 29.80\\ \hline

Deeplabv3+ (Xception)~\cite{chen2018encoder} & 0.6746 ± 0.2746 & 0.5327 ± 0.2788 & 0.7757 ± 0.2967 & 0.6296 ± 0.2699 & 41.25M & 39.62\\\hline

Deeplabv3+ (Mobilenet)~\cite{chen2018encoder} & 0.6474 ± 0.2634 & 0.5098 ± 0.2653 & 0.6632 ± 0.2611 & 0.6878 ± 0.2916 & 2.10M & 78.63\\ \hline

ColonSegNet~\cite{jha2021real} & 0.6324 ± 0.2772 & 0.5183 ± 0.2830 & 0.6112 ± 0.3058  & 0.7897 ± 0.2731 & 5.01M & 7.92 \\ \hline

ResUNet++ + CRF~\cite{jha2021comprehensive}& 0.4200 ± 0.3150 & 0.6096 ± 0.2770 & 0.3782 ± 0.3132 & 0.6711 ± 0.4004 & 4.02M & 46.98  \\ \hline

PraNet~\cite{fan2020pranet}  & 0.7293 ± 0.3004 & 0.6262 ± 0.3128 & \textbf{0.8007} ± 0.2675 & 0.7623 ± 0.3243 & 32.54M & 49.61  \\ \hline
UACANet-S~\cite{kim2021uacanet} & 0.6945 ± 0.2634 & 0.5894 ± 0.2911 & 0.7692 ± 0.2424 & 0.7377 ± 0.3119 & 26.90M & 32.89 \\ \hline
UACANet-L~\cite{kim2021uacanet}& 0.7312 ± 0.2724  & 0.6383 ± 0.2961 & 0.7417 ± 0.2803 & 0.8314 ± 0.2627 & 69.15M & 32.73 \\ \hline

MSRF-Net (Ours) &  \textbf{0.7575} ± 0.2643 & \textbf{0.6337} ± 0.2815 &0.7197 ± 0.2775 &0.8414 ± 0.2731 & 18.38M & 16.24\\ \hline   
\bottomrule
\end{tabular}
\vspace{-7mm}
\end{table*}

\begin{table*}
\centering
\caption{Ablation study of \sysname on the Kvasir-SEG}
\label{table:ablationstudy}
\scriptsize
\begin{center}
{
	\begin{tabular}{@{}l|r|r|r|r|r|r@{}} 
		\toprule
		 \textbf{Experiment description} & \textbf{DSC} & \textbf{mIoU} & \textbf{Recall} & \textbf{Precision} & \textbf{Parameters} & \textbf{FPS} \\ 
		 \hline
		 \hline
		 \sysname(ours) & \textbf{0.9217} ±  0.1685 & \textbf{0.8914} ± 0.1938 & 0.9198 ± 0.1919 & \textbf{0.9666} ± 0.1379 & 18.38M & 16.24\\ \hline
		 Without sub-network &0.8771 ± 0.2062 &0.8103 ± 0.2400 & 0.8911 ±  0.1742 & 0.8993 ± 0.2172 &  \textbf{2.66M} &  \textbf{32.64} \\ \hline
		 Sub-network without scaling &0.9137 ± 0.1772 &0.8898 ± 0.2013 &0.9625 ± 0.1932 &0.9218 ± 0.1484 & 18.38M & 8.92 \\ \hline
		 Sub-network without DSDF (across 2,3 scale) & 0.9013 ± 0.2111 & 0.8782 ± 0.2329 & 0.9460 ± 0.2098 & 0.9246 ± 0.2068 & 17.24M & 17.28  \\\hline
		 Subset of the sub-network & 0.8986 ± 0.1997 & 0.8570 ±  0.1262 & 0.9228 ± 0.2879 & 0.9232 ± 0.2666 & 8.67M & 13.10  \\ \hline
		 Without deep supervision & 0.8988 ± 0.2060 & 0.8449 ± 0.2313 & 0.9053 ± 0.1795 & 0.9267 ± 0.2152 & 18.38M & 16.58 \\ \hline
		 Without decoder block & 0.9067 ± 0.1834 & 0.8691 ± 0.2080 & 0.9143 ±  0.1875 & 0.9461 ± 0.1807 &17.41M & 16.90 \\ \hline
		 Without shape stream & 0.9194 ± 0.1779 & 0.8907 ± 0.1984 & \textbf{0.9700} ± 0.1976 & 0.9159 ± 0.1348 & 18.37M & 17.19\\
		 		 
		 \hline
		 MSRF-Net with \lcomb = \ldcs & 0.8861 ± 0.2192 & 0.8446 ± 0.2389 & 0.9139 ± 0.2078 & 0.9176 ± 0.2003 & 18.38M & 16.18\\ \hline
		 MSRF-Net with \lcomb = \lbce & 0.9059 ± 0.1859 & 0.8677 ± 0.2116 & 0.9446 ± 0.1938 & 0.9143 ± 0.17000 & 18.38M & 16.11\\ \hline
		 \bottomrule
\vspace{-10mm}		 
	\end{tabular}}
\end{center}
\end{table*}

\section{Discussion}
%
Multi-scale fusion methodologies have been studied previously, however, there are some disadvantages. For example, U-Net~\cite{ronneberger2015u} uses skip-connections for feature fusion, but the resulting combination of features suffer from semantic gap since it combines low level features of the encoder and high level features of the decoder. Similarly, U-Net++~\cite{zhou2019unet++} performs low to high feature fusion to overcome this problem, but high to low feature fusion remains lacking. Pyramid features are fused in Deeplabv3+~\cite{chen2018encoder} while without maintaining the high resolution representations. Similar to our approach, HR-Net builds upon the multi-scale feature fusion process by adding repeated feature fusion while keeping high resolution representation, however, their fusion modules consist of a larger number of trainable parameters and informative low level features are also lost during the segmentation process~\cite{xu2021hrcnet}. 
The disadvantages stated above can result in Deeplabv3+~\cite{chen2018encoder} and HRNetV2~\cite{wang2020deep} to perform considerably worse on the 2018 Data Science Bowl challenge where finer segmentation maps were required for a high \ac{DSC} score. The results on Kavisir-SEG, CVC-ClinicDB, and ISIC-2018 also show similar performance gaps between proposed and other multi-scale fusion methods (see Table I - IV). 

\begin{figure}[!t]
    \centering
    \includegraphics[width=0.45\textwidth,height=0.2\textwidth]{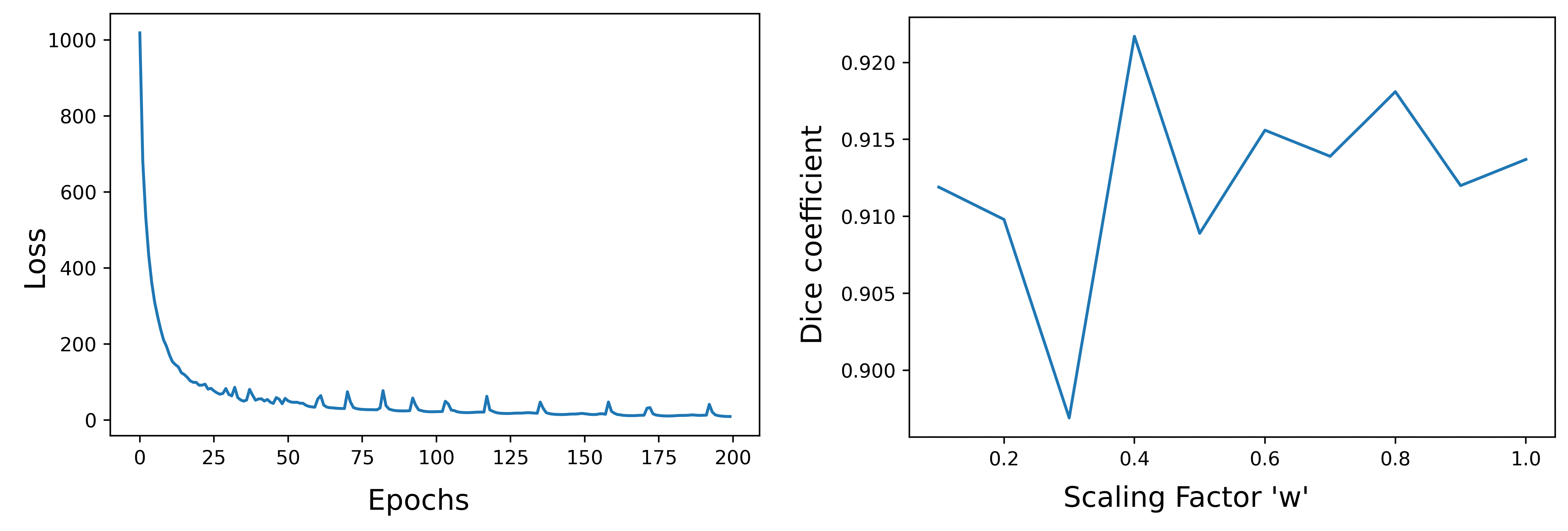}
        \caption{ a) Plot of training loss versus number of epochs and b) Variation in \ac{DSC} with respect to hyper parameter $w$.}
    \label{fig:hyperparamw}
    \vspace{-5mm}
\end{figure}

The proposed \sysname uses \ac{DSDF} blocks (arranged as described in Algorithm~\ref{alg:algo1}) to attain global multi-scale fusion while increasing the frequency of multi-scale fusion operations and reporting a lower computational complexity as compared to HRNetV2~\cite{wang2020deep}. The \ac{DSDF} blocks itself allow effective feature fusion between high- and low resolution scales by continuous feature exchange across different scales. Additionally, its residual structure permits the relevant high- and low-level features to be deftly propagated enabling the proposed \sysname to effectively capture the variability in size, shape and structure of the region of interest. We can observe that the residual densely connected nature of the \ac{DSDF} blocks and its subsequent arrangement allows our proposed \sysname to achieve highest \ac{DSC} of 0.9217 and \ac{mIoU} of 0.8914,  on the Kvasir-SEG (see Table~\ref{tab:result1}). Similarly, we report the highest values for \ac{DSC}, \ac{mIoU} and recall of 0.9420, 0.9043 and 0.9567, respectively, on CVC-ClinicDB (see Table~\ref{tab:result2}). The ability of our \sysname to recognize smaller and finer cell structures in 2018 Data Science Bowl is evident in Table~\ref{tab:result3}, where we report the best \ac{DSC} of 0.9224. Additionally, we  report best \ac{mIoU} and recall on the ISIC-2018 skin lesion dataset. Our result is competitive to DoubleUNet in terms of \ac{DSC}. We present training loss of \sysname (Kvasir-SEG) with respect to the number of epochs elapsed in Figure~\ref{fig:hyperparamw}(a). We can see that the model starts converging from epoch number 75 steadily.

In practical clinical environments, the performance of deep learning based segmentation methods decreases due to differences in the imaging protocols and patient variability. The  models  which  are  able  to  generalize  across multi-center dataset are more desirable in a clinical setting~\cite{ali2021polypgen}. \sysname achieves the highest \ac{DSC} of 0.7921 when trained on Kvasir-SEG and tested on CVC-ClinicDB (see Table~\ref{tab:generalizationkvasir}). Similarly, \sysname achieves highest \ac{DSC} of 0.7575  and \ac{mIoU} of  and 0.6337, when trained on CVC-ClinicDB and tested on Kvasir-SEG (see Table~\ref{tab:generalizationcvc}). HRNetV2-W48~\cite{wang2020deep} was competitive to our method. The above results suggest that our proposed MSRFNet is more generalizable. This can be evidently due to our multi-scale fusion that exploits the feature at different scales, preserving some class representative features.

We performed an ablation study (see Table~\ref{table:ablationstudy}) to demonstrate that the combination of relevant high- and low-level multi-scale features obtained by the MSRF sub-network is instrumental in recognizing the shape or boundaries of the target object that can boost the segmentation performance. To verify the contribution of the \ac{MSRF} sub-network, we disable the entire  \ac{MSRF} sub-network from the full network while keeping each component of the network intact and train the model. Table~\ref{table:ablationstudy} shows that when the \ac{MSRF} sub-network is removed from the proposed \sysname, the \ac{DSC} drops by 4.46\%. This performance degradation illustrates that each \ac{MSRF} sub-network contributes to the network. The combination of high- and low-level resolution feature representations of varying receptive fields extracted from the \ac{MSRF} sub-network contribute significantly towards improving the model's performance. We also ablated if multi-scale fusion was suitable for the entire network. Sub-Network without \ac{DSDF} refers to the removal of \ac{DSDF} with 2nd and 3rd scale inputs (also see Figure~\ref{fig:smfrdbblock}(b), where middle \ac{DSDF} blocks represent them, i.e., layer three and layer five are removed). Table~\ref{table:ablationstudy} shows the result when global multi-scale fusion is absent from the network. As a result, we observe a 2.04\% performance drop in \ac{DSC}. Therefore, it is noticeable that the multi-scale fusion used in the \ac{MSRF} sub-network improves performance.
To study the impact of the number of \ac{DSDF} blocks on the segmentation performance, we reduced the number of \ac{DSDF} layers from six (ours) to three, i.e., only red rectangular block in Figure~\ref{fig:smfrdbblock}(b) is used. Even though this enables us to exchange global multi-scale feature representations, our results in Table~\ref{table:ablationstudy} show that reducing the number of \ac{DSDF} blocks decreases the \ac{DSC} by 2.31\% .

\begin{figure}[t!]
\centering
\includegraphics[width=0.35\textwidth]{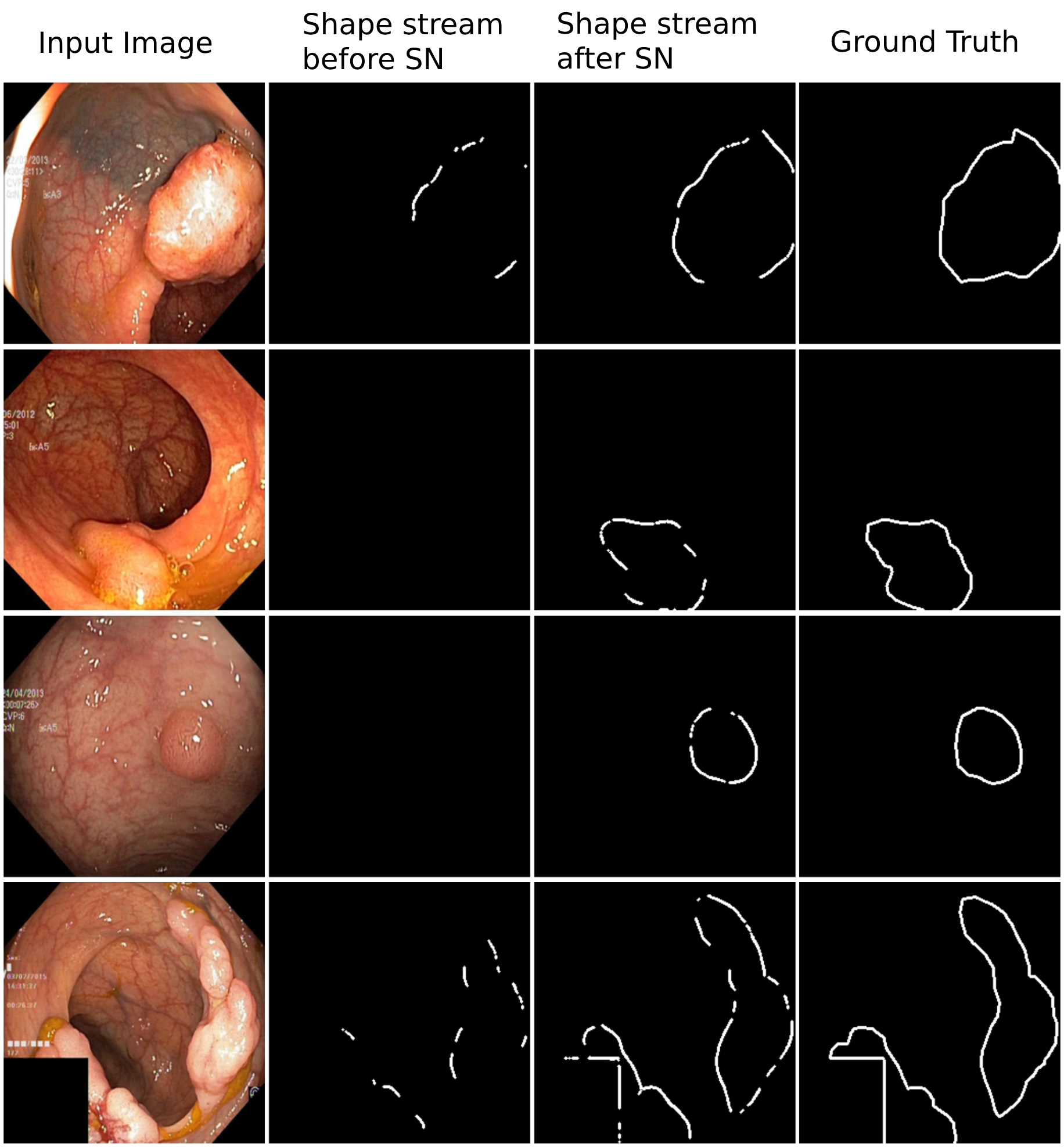}
\caption{Qualitative results showing polyp contours when shape stream is used before \ac{MSRF} sub-network and after the \ac{MSRF} sub-network in the  MSRF-Net}
\label{fig:shapestream}
\vspace{-5mm}
\end{figure}

The sub-network without scaling in Table~\ref{table:ablationstudy} demonstrates the influence of scaling factor $w$ in the network (see Equation~\ref{eq:5}). For this experiment, we did not scale the output of \ac{DSDF} by a constant while adding to the block's input. Drop of 0.80\% in \ac{DSC} was observed when the features were not scaled. Furthermore, our empirical experiments (see Figure~\ref{fig:hyperparamw}(b) using different scaling values of $w$ introduced in Equation~\ref{eq:5}) demonstrate our optimal choice of $w$ to be $0.4$.

We design a variant model where, the \ac{MSRF} sub-network is placed after the shape-stream in the \sysname. Here,  we keep the number of parameters same for both the models (i.e., \sysname and variant model) to analyze the impact of \ac{MSRF} sub-network on the shape stream.  The qualitative results (see Figure~\ref{fig:shapestream}) show that the \sysname can define more precise and more spatially accurate boundaries than the variant model. The variant model fails to recognize the boundaries of the target structure as it is deprived of the multi-scale features extracted by the \ac{MSRF} sub-network. This validates our choice of putting the \ac{MSRF} sub-network before the shape stream block. Only a minor drop of 0.23\% in \ac{DSC} is seen when no shape stream is applied and it still outperforms most \ac{SOTA} methods.
%

We also investigated the impact of our triple attention block by disabling the mechanism prior to training the \sysname.
We disable deep supervision in another experiment, while training \sysname. Both of these experiments showed performance drop compared to our proposed \sysname (1.50\% drop in former and 2.29\% drop in latter on \ac{DSC} metric). We also evaluate the impact of the combination of \lbce and \ldcs used in \lcomb (see Section~\ref{section:loss_comp}). For this, we trained the \sysname with \lcomb = \ldcs and then with \lcomb = \lbce.
When \lcomb = \ldcs + \lbce, we obtained an increase of 3.56\% in \ac{DSC}, 4.68\% in \ac{mIoU}, 0.59\% in recall and 4.90\% in precision as compared to the \lcomb = \ldcs setting. Similar trend was observed when \lcomb was equal to \lbce (see Table~\ref{table:ablationstudy}).

%
%
%

 \sysname clearly shows the strength of fusing low- and high-resolution features through DSDF blocks and MSRF sub-network. Alongside, complementary inclusion of scaling factor, deep supervision in the encoder block and triple attention in the decoder block showed further improvements. In Figure~\ref{fig:suboptimal}, we show the qualitative results for the sub-optimal cases. The qualitative results show poor performance for oblique samples in polyp datasets. Similarly, the model also failed for extremely low contrast images with 2018 DSB and scattered similar patches in ISIC 2018. 

\vspace{-3mm}

\section{Conclusion}

In this paper, we proposed the \sysname architecture for medical image segmentation that takes advantage of multi-scale resolution features passed through a sequence of \ac{DSDF} blocks. Such densely connected residual blocks with dual-scale feature exchange enable efficient feature extraction with varying receptive fields. Additionally, we have also shown that the features from \ac{DSDF} blocks are better suited to capture a target object's entire shape boundaries, even for objects with variable sizes. Our experiments revealed that \sysname outperformed several \ac{SOTA} methods on four independent biomedical datasets. Our investigation using cross-datasets testing to evaluate the generalizability of the \sysname confirmed that our model can produce competitive results in such scenarios. We also identified some challenges of the proposed method, such as that the model fails when extremely low contrast images are part of the data. For future work, we plan to investigate the identified challenges further and adjust the design of the network to address the challenging cases. 

\section*{Acknowledgment}
D. Jha is funded by the PRIVATON project (\#263248) which is funded by Research Council of Norway (RCN). S. Ali is supported by the National Institute for Health Research (NIHR) Oxford Biomedical Research Centre (BRC). The computations in this paper were performed on equipment provided by the Experimental Infrastructure for Exploration of Exascale Computing (eX³), which is financially supported by the Research Council of Norway under contract 270053.

\vspace{-2mm}

\bibliographystyle{IEEEtran}
\vspace{-2mm}
\bibliography{references} 
\end{document}